\documentclass[a4paper]{IEEEtran}
\usepackage[a4paper, total={7in, 9.8in}]{geometry}
\usepackage[boxruled]{algorithm2e}
\usepackage{enumerate}
\usepackage{multirow,array}
\usepackage{cite}
\usepackage{graphicx}
\usepackage{psfrag}
\usepackage{caption}
\usepackage{subcaption}
\usepackage{url}
\usepackage{amsmath}
\usepackage{amsthm}
\usepackage{array}
\usepackage{algorithm2e}
\usepackage{amssymb}
\usepackage{amsfonts}
\usepackage{float}

\makeatletter
\renewcommand{\boxed}[1]{\text{\fboxsep=.2em\fbox{\m@th$\displaystyle#1$}}}
\makeatother
\newcommand{\C}{\cal}
\newcommand{\Lbarsq}{\overline{{\cal L}^2}}
\newcommand{\Lsq}{{\cal L}^2}
\newcommand{\B}{\boldsymbol}
\newcommand{\fq}{{\mathbb F}_q}
\newcommand{\gbinom}[2]{\begin{bmatrix}#1\\#2\end{bmatrix}_q}
\newtheorem{proposition}{Proposition}
\newtheorem*{conjecture*}{Conjecture}
\newtheorem{lemma}{Lemma}

\newtheorem{remark}{Remark}
\newtheorem{definition}{Definition}
\newtheorem{theorem}{Theorem}
\newtheorem{example}{Example}

\title{Coded Caching via Line Graphs of Bipartite Graphs}
\begin{document}
\author{
\IEEEauthorblockN{Prasad Krishnan\\}
\IEEEauthorblockA{
International Institute of Information Technology, Hyderabad.
\\
Email: prasad.krishnan@iiit.ac.in
}}
\date{\today}
\maketitle
\thispagestyle{empty}	
\pagestyle{empty}
\begin{abstract}
We present a coded caching framework using line graphs of bipartite graphs. A clique cover of the line graph describes the uncached subfiles at users. A clique cover of the complement of the square of the line graph gives a transmission scheme that satisfies user demands. We then define a specific class of such caching line graphs, for which the subpacketization, rate, and uncached fraction of the coded caching problem can be captured via its graph theoretic parameters. We present a construction of such caching line graphs using projective geometry. The presented scheme has a rate bounded from above by a constant with subpacketization level $q^{O((log_qK)^2)}$ and uncached fraction $\Theta(\frac{1}{\sqrt{K}})$, where $K$ is the number of users and $q$ is a prime power. We also present a subpacketization-dependent lower bound on the rate of coded caching schemes for a given broadcast setup. 
\end{abstract}
\section{Introduction}
Wireless data traffic has been growing tremendously in the last decade and is expected to do so in the future, and video in particular is expected to comprise of more than 50\% of the total traffic \cite{Cis}. Caching has been in vogue to lay off the traffic during peak times in the network by storing part of the information demanded by users (clients) in local storage known as \textit{caches}. In this way, during the peak hours, the server can transmit only the non-cached information thus reducing the traffic. For instance, consider the setting taken in \cite{MaN}, consisting of a single-server error-free broadcast channel with $K$ clients(users), $N$ files at the server  (each file comprised of $F$ subfiles, where $F$ is known as the \textit{subpacketization } parameter), with each client capable of caching $MF$ subfiles. By populating the cache during the caching phase when the demand requests are not present, traditional caching can achieve a rate $R$ equal to $K(1-\frac{M}{N})$ during the delivery phase when the network is required to satisfy the demands (the rate $R$ is defined as fraction of the number of transmitted packets (each of size equal to a subfile) to $F$.

 The paradigm of Coded Caching, introduced in \cite{MaN}, was based on the idea of transmitting coded packets during the peak times to further reduce the network usage. In contrast to the uncoded caching scenario which required a rate $K(1-\frac{M}{N})$, the coded caching scheme shown in \cite{MaN} achieves a rate $\frac{K(1-\frac{M}{N})}{1+\frac{MK}{N}}$, which is constant as $K\rightarrow \infty$ for constant $\frac{M}{N}$. This is achieved by designing both the caching phase and the delivery phase carefully. The tremendous rate advantage shown in \cite{MaN} were shown in other settings also, for instance \cite{JCM,PMN}. The scheme in \cite{MaN} is also shown to be order optimal, i.e., within a constant multiple of the optimal rate for the same set of parameters. The scheme of \cite{MaN} is achieved by dividing each file into $F=\binom{K}{\frac{MK}{N}}$ subfiles, and caching them appropriately in the client caches. It was noticed in \cite{SJTLD} that the subpacketization required is exponential for constant $\frac{M}{N}$ as $K$ grows large (as $\binom{K}{Kp}\approx 2^{KH(p)}$ for constant $0<p<1$, where $H(p)$ is the binary entropy).  Since then a number of papers \cite{SJTLD,YTCC,YCTC,SZG,STD,TaR,CYTJ,CJYT,SLB} have presented new schemes for coded caching which uses smaller subpacketization at the cost of having increased rate or cache requirement compared to \cite{MaN}. Table \ref{tab1} lists the relevant known results in this context (the references and the techniques used are shown in the first column). The second column lists the uncached fraction of any file (a fraction $\frac{M}{N}$ of each file is cached by a user). Many of the schemes presented in Table \ref{tab1} require exponential subpacketization (in $K$, for large $K$), as shown in the fourth column of Table \ref{tab1} to achieve a constant rate (shown in the last column). The subpacketization of particular schemes of \cite{SZG,CJYT} have been shown to be sub-exponential, while some schemes of \cite{CYTJ} have subpacketization that is linear or polynomial (in $K$) at the cost of either requiring larger cache $M$ or larger rate compared to \cite{MaN}. Interestingly, a linear subpacketization scheme ($F=K$) was shown in \cite{STD} using a graph theoretic construction with near constant rate and small memory requirement. However the construction in \cite{STD} holds for very large values of $K$ only. In \cite{SZG}, it was shown that subpacketization linear in $K$ is impossible if we require constant rate. In \cite{SLB}, the authors consider caching schemes without file splitting, i.e., the scenario when $F=1$. 
\begin{table*}
\scriptsize
\begin{tabular}{|c|c|c|c|c|}
\hline
Scheme& $(1-\frac{M}{N})$ & Number of Users $K$ &   $F$  & Rate $R$\\
& & &  large $K$ and constant $\frac{M}{N}$ & \\
\hline
Ali-Niesen \cite{MaN} & $(1-\frac{M}{N})$ for $M<N$    & any $K$ & $c_1e^{Kd_1}$ & $\frac{K-t}{1+t}=\frac{K(1-\frac{M}{N})}{\frac{MK}{N}+1}$\\
& such that $\frac{MK}{N}\in {\mathbb Z}_{\geq 0}$  & &(for large  $K$) &\\
\hline
Ali-Niesen Scheme with & Same as \cite{MaN} & $K$ & $O(c_4e^{d_4K})$ ($c_2,c_3<d_4$) & $\frac{K}{g+1}\left(1-\frac{1}{\lceil\frac{N}{M}\rceil}\right)$, where \\
Grouping \cite{SJTLD} &&&(for large  $K$)&$g\in {\mathbb Z}$ such that $\frac{K}{g\lceil\frac{N}{M}\rceil}\in{\mathbb Z}$.\\
\hline
Yan et al \cite{YCTC} (PDAs) & $1-\frac{1}{q}$ or $\frac{1}{q}$  & Any $K$ &$c_2e^{Kd_2}$ ($d_2<c_2$, for large $K$) & $\frac{K(1-\frac{M}{N})}{\frac{MK}{N}}$\\
\hline
Shanguan et al \cite{SZG} &	 & &For large $K$ & $R\approx (2q-1)^2$, such that $q=\frac{\lambda}{2}$,  \\
(PDAs based on hypergraphs)& $1-\frac{1}{q}$ or $\frac{1}{q}$ & Specific choices & $c_3e^{\sqrt{K}d_3}$ &where $\lambda$ is such that $\frac{M}{N}=\frac{2\lambda-1}{\lambda^2}$\\ 
\hline
Yan et al \cite{YTCC} (for integers $0<a,b<m$  &&&&\\
and $\lambda <min\left\{a,b\right\}$ based on & $1-\frac{\binom{a}{\lambda}\binom{m-a}{b-\lambda}}{\binom{m}{a}}$ & $\binom{m}{a}$ & $\binom{m}{b}$ &$\frac{\binom{m}{a+b-2\lambda}\binom{a+b-2\lambda}{a-\lambda}}{\binom{m}{b}}$\\
strong edge coloring of bipartite graph) & && &\\
\hline
Tang et al \cite{TaR} based & & & For large $K$, $c_5e^{Kd_5}$, exponent   & \\
on resolvable designs & $1-\frac{1}{q}$ or $\frac{1}{q}$ & $nq$  & similar to \cite{YTCC} and \cite{SZG}  & $\frac{K(1-\frac{M}{N})}{log(F)-c}$\\
&&(for some constant $q$)& (some schemes) but less than & (for some constant $c$)\\
&&&  some schemes of \cite{SZG}&\\
\hline
Scheme from \cite{STD} based on & $\leq (1-K^{-\epsilon})$ & $K$ &  &\\
induced matchings of a & (where $\epsilon=k_1\delta e^{-\frac{k_2}{\delta}}$, & (necessarily & $K$ 	&$K^\delta$ \\
Rusza Szemeredi graph & for $\delta$ as in last column) &  large) & & (some small $\delta$)\\
\hline
PDA scheme $P_1$ from & For integers $k,t$  &&&\\
Cheng et al \cite{CYTJ}  & $\frac{t+1}{\binom{k}{t}}$ & $\binom{k}{t+1}$ & $\binom{k}{t}$ & $\frac{k}{\binom{k}{t}}$\\
\hline
PDA scheme $P_2$ from & For integers $k,t$ &&&\\
Cheng et al \cite{CYTJ}  & $1-\frac{t}{k}$ & $\binom{k}{t}$ & $k$ & $\frac{\binom{k}{t+1}}{k}$\\
\hline
Two PDA Schemes from \cite{CJYT} & For integers $z,q,t$ & $\binom{m}{t}q^t$ and & &$\left((q-z)/{\lfloor \frac{q-1}{q-z}\rfloor}\right)^t$ \\
&$\left(\frac{(q-z)}{q}\right)^t$ and $1-\frac{z}{q}$ & $(m+1)q$ &  $O\left(q^{\frac{tK^{\frac{1}{t}}}{q}}\right)$ & and $(q-z)/{\lfloor \frac{q-1}{q-z}\rfloor}$\\
\hline
\end{tabular}
\caption{Known results}
\label{tab1}
\end{table*}

The contributions and organization of this work are as follows. After reviewing the work of \cite{YTCC} in Section \ref{bipartite}, we prove a lower bound on the peak delivery rates of coded caching schemes using the properties of the associated bipartite graph (Section \ref{lowerbound}). We then map the problem of finding a valid transmission scheme corresponding to a  bipartite caching scheme to a clique cover problem of a graph derived from the line graph of the bipartite graph (Section \ref{sec4linegraphs}). In Section \ref{sec4linegraphs}, we also show that the existence of a class of such line graphs of bipartite graphs implies the existence of coded caching schemes for which there is a nice characterization of the rate, uncached fraction, and subpacketization. We then give a coded caching scheme using a construction of such caching line graphs based on projective geometries over finite fields (Section \ref{ourscheme}). Analyzing this scheme in Section \ref{analysis}, we get to add results to Table \ref{tab1}, as shown in Table \ref{tab2}. The first row of Table \ref{tab2} lists the actual parameters of our scheme. The other two rows indicate asymptotic results as $K\rightarrow \infty$ (with constant field size $q$) respectively. We note that the last row shows a constant rate with sub-exponential packetization achieved by the scheme in this paper. We conclude the paper with a short discussion in Section \ref{conclusion}. 

\begin{table}
\centering
\begin{tabular}{|c|c|c|c|}
\hline
$1-\frac{M}{N}$&$K$&$F$&$R$\\
\hline
\multicolumn{4}{|c|}{\textbf{For non-negative integers $k,m,t$ with $1\leq m+t\leq k$}}\\
\hline
$\frac{\gbinom{m+t}{t}}{\gbinom{k}{t}}$&$\gbinom{k}{t}$&$\gbinom{k}{m+t}$&$\frac{\gbinom{m+t}{t}}{\gbinom{k-m}{t}}$\\
\hline
\multicolumn{4}{|c|}{\textbf{Limiting behaviors as $k$ grows, for constants $t, k-m,q : $}}\\
\hline
$\geq q^{(m+t-k-1)t}$&$\gbinom{k}{t}$&$\leq K^{\frac{k-t-m+1}{t}}$ &$\leq\frac{K}{q^{2(k-m-t-1)t}}$\\
~~(constant) & & $(O(poly(K))$&$( \Theta(K) )$\\
\hline
\multicolumn{4}{|c|}{\textbf{Limiting behaviors as $k$ grows, for constants $t, k-2m, q : $}}\\
\hline
$\Theta(\frac{1}{\sqrt{K}})$&$\gbinom{k}{t}$&$q^{O((log_qK)^2)}$ &$\leq q^{(2m+t-k+1)t}$\\
&&&(constant)\\
\hline
\end{tabular}
\vspace{0.2cm}\caption{Parameters of Scheme in Section \ref{ourscheme}. (proofs of last two rows in Section \ref{analysis})}
\label{tab2}
\end{table}

\textit{Notations and Terminology: }
For a positive integer $n$, we denote by $[n]$ the set $\{1,\hdots,n\}$. We recall only minimal facts regarding graph theory. For other standard definitions, the reader is referred to \cite{Die}. A graph $G$ consists of a set $V(G)$ of vertices and a set $E(G)\subset \left\{\{u,v\}:u,v\in V(G)\right\}$ of edges. 
For a subset $A$ of vertices of graph $G$, we denote ${\mathcal N}(A)$ as the set of adjacent vertices of $A$. 
 A \textit{bipartite graph} $B$ is one whose edges can be visualized as being between two subsets of a partition of the vertex set (called \textit{left} and \textit{right} vertices of $B$).  
A subset $S\subseteq V$ is called a \textit{clique} of $G$ if all vertices in $S$ are adjacent to each other (we assume vertices to be cliques of size $1$). A $b$-\textit{clique-cover} of $G$ is a collection of cliques $D_i:i=1,\hdots,b$ such that $\cup_i D_i=V.$ 

\section{Bipartite Graph based Coded Caching and Delivery based on \cite{YTCC}}
\label{bipartite}
Let $\cal K$ be the set of users(clients) ($|{\cal K}|=K$) in a system consisting of one server having files $\{W_i:i\in[N]\}$ connected to the clients via a error-free broadcast channel. Let $F$ be the subpacketization level, i.e. each file is composed of $F$ subfiles, each taking values according to a uniform distribution from some finite abelian group ${\cal A}$. The subfiles of file $W_i$ are denoted as $W_{i,f}:f\in {\cal F}$ for some set ${\cal F}$ of size $F$. Let $MF$ denote the number of subfiles that can be stored in the cache of any user. A \textit{coded caching scheme }consists of two subschemes (as in \cite{MaN}), a \textit{caching scheme} according to which subfiles of the files are placed in the user caches during periods when the traffic is low, and a \textit{transmission scheme} that consists of  broadcast transmissions from the server satisfying the demands of the clients appearing during the demand phase. We assume \textit{symmetric caching} throughout the paper, i.e., the caches at the users are populated in such a way that if user $k\in {\cal K}$ caches the subfile $f\in {\cal F}$ of any file, then it caches the subfile $f\in {\cal F}$ of each file. All the schemes presented in \cite{SJTLD,YTCC,YCTC,SZG,STD,TaR,CYTJ,CJYT} employ symmetric caching. We also assume throughout this work that $\frac{MF}{N}$ is an integer which is the number of subfiles of any particular file stored in a user's cache. In the delivery scheme, the transmissions (of size equal to subfiles) must be done so that the demands of the clients are all satisfied. As in \cite{MaN}, the rate $R$ of the coded caching scheme is defined as 
\[
\text{Rate}~R = \small \frac{\text{Number of transmissions in the transmission scheme}}{\text{Number of subfiles in a file}}.
\]

\begin{figure}
  \centering
    \begin{subfigure}[b]{0.2\textwidth}
                \centering
                \includegraphics[height=1.5in]{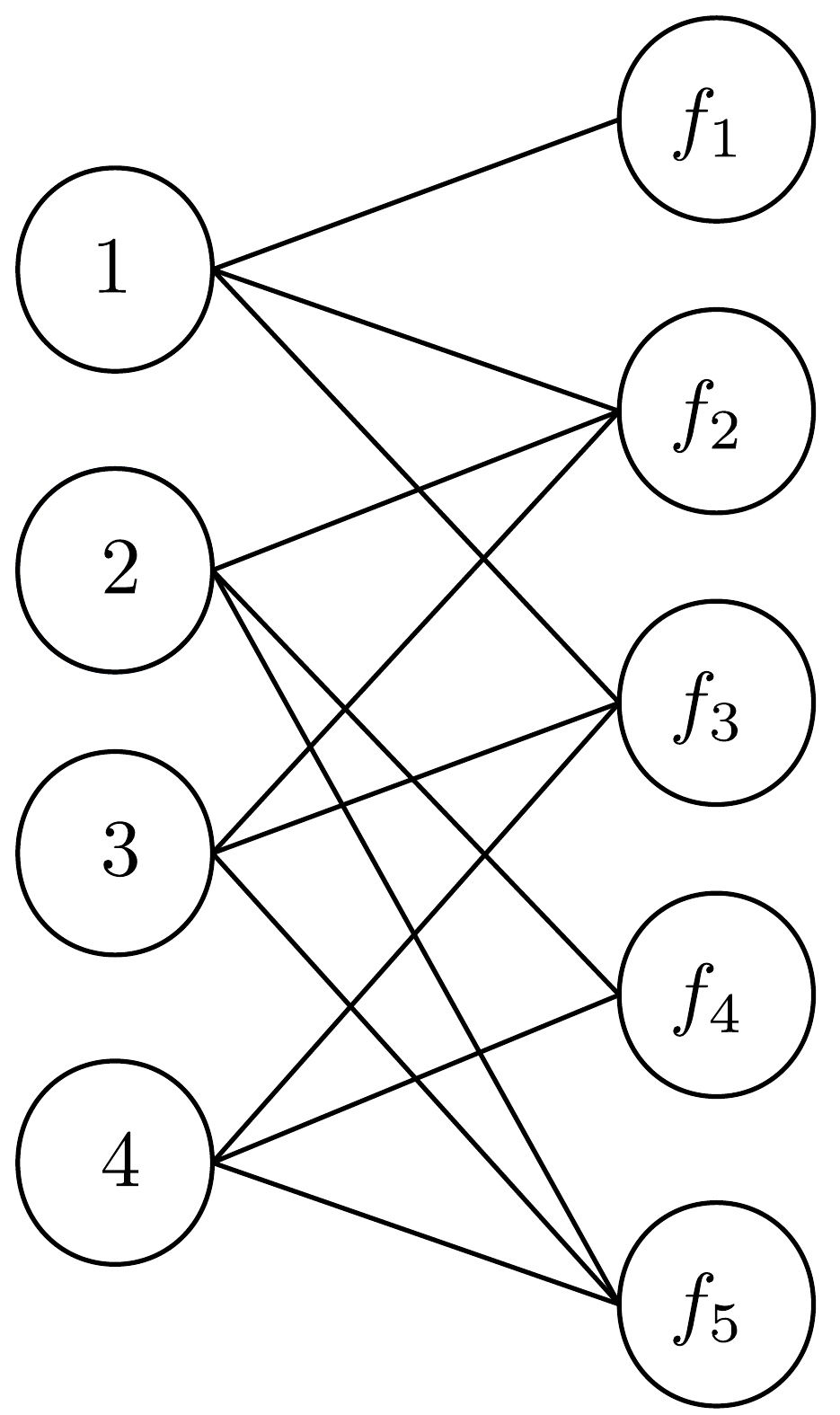}
\label{fig:bipartite}
        \end{subfigure}
        \begin{subfigure}[b]{0.19\textwidth}
                \centering
                \includegraphics[height=1.2in]{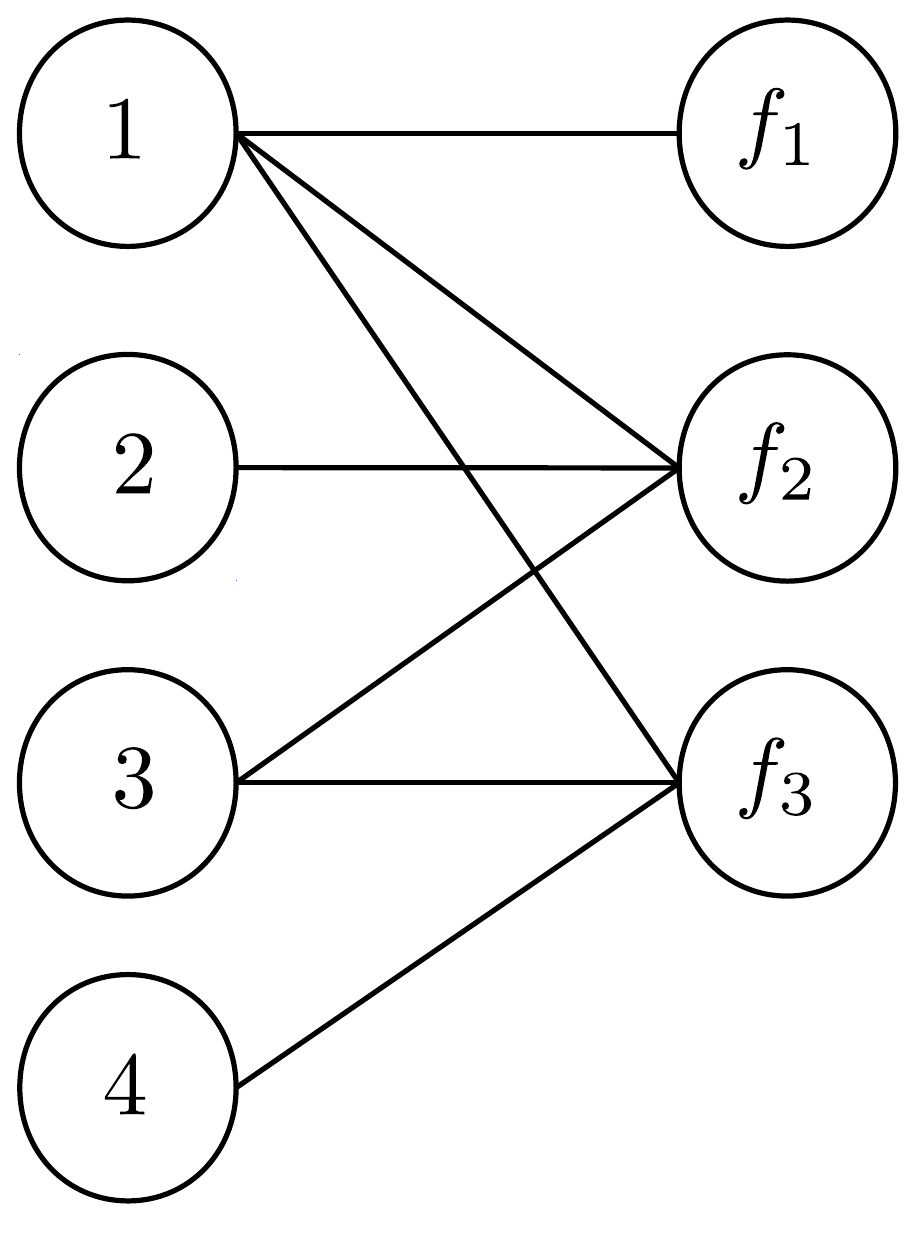}
          \label{fig:bipartiteinduced}      
        \end{subfigure}%
        \caption{The left figure is a bipartite caching scheme with $4$ users and $5$ subfiles with $1-M/N=3/5$. Edges indicate missed subfiles. The right figure shows the subgraph induced by ${\C N}(1)\cup{\C N}({\C N}(1))$.}
        \label{bipartitecaching}
       \end{figure}

We can visualize the symmetric caching scheme (with fully populated caches) using a bipartite graph, following \cite{YTCC}. Consider a bipartite graph $B$ with ${\cal K}$ being the left(user) vertices and the right(subfile) vertices being ${\cal F}$. We then define the edges of the bipartite graph to denote the uncached subfiles of the files, i.e, for $k\in {\cal K}, f\in {\cal F}$, an edge $\{k,f\}\in E(B)$ exists if and only if user $k$ does \textit{not} contain in its cache the subfile $f$ of each file. Clearly, this bipartite graph is left-regular, with $F\left(1-\frac{M}{N}\right)$ being the degree of any user vertex. Indeed any left-regular bipartite graph defines a caching scheme, which we formalize below. 
\begin{definition}[Bipartite Caching Scheme]
Given a bipartite $D$-left-regular graph with $K$ left vertices and $F$ right vertices denoted by $B(K,D,F)$ (or in short, $B$), the symmetric caching scheme defined on $K$ users with subpacketization $F$ with the edges of $B$ indicating the uncached subfiles at the users, is called the $(K,D,F)$ bipartite caching scheme associated with the bipartite graph $B$.
\end{definition}
\begin{remark}
We observe that the bipartite caching scheme associated with the graph $B(K,D,F)$ has the uncached fraction $1-\frac{M}{N}=\frac{D}{F}$.
\end{remark}
Fig. \ref{bipartitecaching} shows a graph describing a $(4,3,5)$ bipartite caching scheme. Note that during the caching phase the user demands are not available. We now look at the transmission phase during which the user $k\in{\cal K}$ demands one file $W_{d_k}$(for some $d_k\in [N]$), as given in \cite{YTCC}. An \textit{induced matching} $\C M$ of a graph $G$ is a matching such that the induced subgraph of the vertices of $\C M$ is $\C M$ itself.   
For an induced matching $\C M$ of $B$ consisting of edges $\left\{\{k_j,f_j\}:j\in[l]\right\}$, consider the associated transmission
\begin{equation}
\label{eqn1}
Y_{\C M}=\sum_{j=1}^l W_{d_{k_j},f_j}
\end{equation}
As ${\C M}$ is an induced matching, $ W_{d_{k_j},f_j}$ is a subfile unavailable but demanded at user $k_j$. By the same reason, each user $k_j$ has all the subfiles in (\ref{eqn1}) in its cache except for $W_{d_{k_j},f_j}$, hence user $k_j$ can decode $W_{d_{k_j},f_j}, \forall j\in[l].$ A $b$-\textit{strong-edge-coloring} of a graph is an assignment of a label (called \textit{colors}) from a finite set ${\cal C}$ of size $b$ to each of its edges such that the  set of all edges of any color (called a \textit{color class}) form an induced matching. Let $\{M_j, j\in[n]\}$ be the set of all induced matchings (color classes) arising from a strong edge coloring of $B$. It is not difficult to see that the transmissions $Y_{M_j}:j\in[n]$ (constructed as in (\ref{eqn1})) corresponding to $M_j:j\in[n]$ satisfies the demands of all the users.  The rate of this transmission scheme $R$ is then $\frac{n}{F}$.

\section{Lower bound on rate of delivery scheme for symmetric caching}
\label{lowerbound}
In this section, we show a bound on the rate of the transmission scheme associated with a $(K,D=F(1-\frac{M}{N}),F)$ bipartite caching scheme associated with $B$. 
As $W_{i,f}$ takes values from ${\cal A}$ with uniform distribution, taking the base of logarithm as $|{\cal A}|$, we have the Shannon entropy of $W_{i,f}$ as $H(W_{i,f})=1,\forall i,f$. Thus $H(W_i)=F$.  For a given $(K,D,F)$ bipartite caching scheme, a rate $R$ is said to be \textit{achievable} if there exists some transmission scheme with rate $R$ that satisfies all client demands. We now prove a lower bound on the infimum $R^*$ of all achievable rates for a given $(K,D,F)$ bipartite caching scheme. 

\begin{theorem}
\label{thmratebound}
Let $k$ be any left-vertex (user vertex) of $B$ and let $H$ be the subgraph of $B$ induced by the vertices ${\cal N}(k)\cup{\cal N}({\cal N}(k)).$ Let $N'=min(N,|{\cal N}({\cal N}(k)|).$ Let $U=\{k_j:j\in[N']\}$ be a subset of $N'$ vertices of ${\cal N}({\cal N}(k))$ taken in some order such that $k_1=k$. For $j\in[N']$, let $\rho_j$ be the set of right vertices (subfiles) in $H$ which are adjacent to $\{k_i:i\in[j]\}.$ Let $R^*$ be the infimum of all achievable rates for the bipartite caching scheme defined by $B$. Then $R^*F\geq \sum_{j=1}^{N'}\rho_j$.

In particular, we must have
\begin{equation}
\label{eqn110}
R^*F\geq min\left( (K+F)\left(1-\frac{M}{N}\right),F\left(1-\frac{M}{N}\right)+N\right)-1.
\end{equation}
\end{theorem}
\begin{IEEEproof}
We are given a valid coded caching scheme with the caching scheme associated with $B$. Let $\boldsymbol{Y}$ denote the set of all transmissions in a valid transmission scheme . As $N'\leq N$, we can assume a demand scenario in which the $N'$ users all demand different files. Let $W_{d_j}$ be the demand of $k_j\in U$ and $Z_j$ be the cache content of user $k_j$. Let $S_j$ denote the set of subfiles of $W_{d_j}$ in the subgraph $H$ missing from users $k_i:i\in[j]$. This corresponds to subfile vertices adjacent to users $k_i:i\in[j]$ in $H$. In our notation, $|S_j|=\rho_j.$ Since $W_{d_j}$s are distinct, thus each subfile in $S_j:j\in[N']$ is distinct. We then follow an idea similar to \cite{YMA}. We construct a virtual receiver which contains an empty cache at first. In the $j^{th}$ step, the cache of this virtual user is populated with all the cache contents of user $j$ except those pertaining to the files demanded by $k_i:i\in[j-1].$ Let $\tilde{Z}_j=Z_j \backslash \{W_{d_i,f}:i<j,\forall f\}.$ Then $\{\tilde{Z}_j:j\in[N']\}$ is the final cache content of this virtual user. By the given transmission scheme, the receivers can decode their demands. Hence, we must have
\begin{equation}
\label{eqn201}
H\left(\{W_{d_j}:j\in [N']\}\mid\{\tilde{Z}_j:j\in[N']\},\boldsymbol{Y}\right)=0,
\end{equation}
as the virtual user must be successively able to decode all the demands of the $N'$ users. Since $RF$ denotes the number of transmissions, we must have the following inequalities. 
\begin{align*}
R^*F&\geq H(\boldsymbol{Y})\\
&\geq I\left(\boldsymbol{Y};\{W_{d_j}:j\in[N']\}\mid\{\tilde{Z}_j:j\in[N']\}\right)\\
&= H\left(\{W_{d_j}:j\in[N']\}\mid\{\tilde{Z}_j:j\in[N']\}\right) ~~(\text{by}~ (\ref{eqn201}))\\
& \geq H\left(\{S_j:j\in[N']\}\right)=\sum_{j=1}^{N'}\rho_j,
\end{align*}
where $I(;)$ denotes the mutual information, and the last inequality is obtained by noting the missing subfiles in $\{\tilde{Z}_j:j\in[N']\}$.

We finally prove (\ref{eqn110}). By a pigeon-holing argument, it is easy to see that there is a subfile vertex having at least $K\left(1-\frac{M}{N}\right)$ adjacent user vertices. Let $k$ be any  user vertex adjacent to such a subfile vertex with $K\left(1-\frac{M}{N}\right)$ adjacent vertices. Consider the subgraph $H$ induced by vertices ${\cal N}(k)\cup {\cal N}({\cal N}(k))$. Note that $|{\cal N}({\cal N}(k))|\geq K\left(1-\frac{M}{N}\right).$ Let $N'=min(N,|{\cal N}({\cal N}(k))|)$. If $N'=N$, consider some subset of ${\cal N}({\cal N}(k))$ containing $k$. Then for any ordering of the $N'$ user vertices starting from $k_1=k$, we have $\sum_{j=1}^{N'}\rho_j\geq F\left(1-\frac{M}{N}\right)+N-1$. If $N'=|{\cal N}({\cal N}(k))|$, then by a similar ordering starting with $k$, we have $\sum_{j=1}^{N'}\rho_j\geq (K+F)\left(1-\frac{M}{N}\right)-1$. Invoking the result in the first part completes the proof.
\end{IEEEproof}
\begin{example}
\label{ex1}
The figure on the right in Fig. \ref{bipartitecaching} shows the subgraph induced by ${\C N}(1)\cup{\C N}({\C N}(1))$ of the bipartite caching graph $B(4,3,5)$ on the left. Assuming the number of files $N\geq 4$, following Theorem \ref{thmratebound},  we can take $\rho_1=3,\rho_2=2,\rho_3=1, \rho_4=0$ (where the user vertices are taken in the order $1,3,2,4$). We then get the minimum rate $R^*\geq \frac{\sum_{i=1}^4\rho_i}{5}=\frac{6}{5}$. 
\end{example}
\section{Line Graphs of Bipartite Graphs and Caching}
\label{sec4linegraphs}
In this section, we shall map the coded caching problem to the line graph of the bipartite caching graph $B$ described in the previous section. The line graph ${\C L}(G)$ of a graph $G$ is a 	graph in which the vertex set $V({\C L}(G))$ is the edge set $E(G)$ of $G$, and two vertices of $V({\C L}(G))$ are adjacent if and only if they share a common vertex in $G$. The square of a graph $G$ is a graph $G^2$ having $V(G^2)=V(G)$, and an edge $\{u,v\}\in E(G^2)$ if and only if $\{u,v\}\in E(G)$ or there exists some $v_1\in V(G)$ such that $\{u,v_1\},\{v_1,v\}\in E(G)$. The following result is folklore and easy to prove.
\begin{lemma}
\label{coloringstrongedge}
There exists a $b$-clique-cover for $\Lbarsq$ if and only if there exists a $b$-strong-edge-coloring for $G$ , with the cliques in the clique cover of $\Lbarsq$ corresponding to the color classes (induced matchings) arising from the strong edge coloring of $G$.
\end{lemma}

\begin{figure}
  \centering
                \includegraphics[height=2.5in]{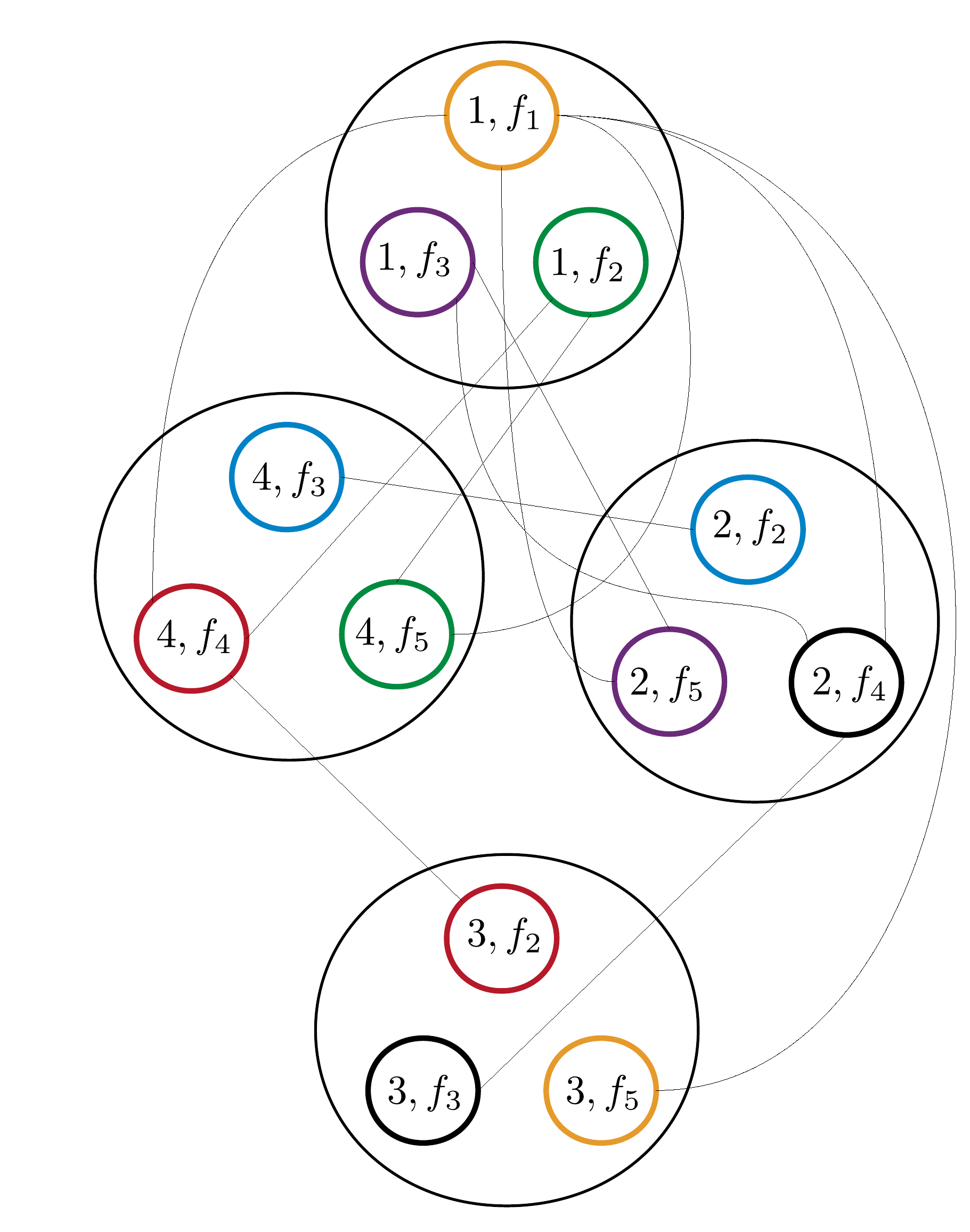}
\caption{The graph $\Lbarsq$ corresponding to the bipartite graph in Fig. \ref{bipartitecaching}. The same-coloured vertices correspond to cliques.}
\label{fig:complementlinesq}
       \end{figure}

By Lemma \ref{coloringstrongedge} and Section \ref{bipartite}, a valid transmission scheme corresponding to the caching scheme associated with $B$ can be obtained by obtaining a clique cover for $\overline{{\C L}^2(B)}$. From the arguments in Section \ref{bipartite}, such a transmission scheme will involve one transmission per each clique in a clique cover of $\Lbarsq$. Fig. \ref{fig:complementlinesq} shows the graph $\Lbarsq$ for the line graph of the bipartite graph shown in Fig. \ref{bipartitecaching}. A clique cover consisting of $6$ cliques is also shown, each containing $2$ vertices. Thus the number of transmissions is $6$, and the rate is $\frac{6}{5}$, which is optimal for this graph as shown in Example \ref{ex1}. 

It turns out that the line graph of the left-regular bipartite graph $B$ is highly structured, and any such structured graph will serve as a line graph of such a bipartite graph.
\begin{proposition}
\label{linegraph}
A graph ${\C L}$ containing $KD$ vertices is the line graph of a $D$-left-regular bipartite graph $B(K,D,F)$ if and only if the following conditions are satisfied.
\begin{itemize}
\item[(C1)] The vertices of $\C L$ can be partitioned into $K$ disjoint cliques containing $D$ vertices each. We denote these cliques by ${\C U}_{k}:k\in[K]$ and call them as the \underline{user-cliques}. We label the vertices of ${\C U}_{k}$ as $e_{k,i}: i\in[D].$  
\item[(C2)] Consider distinct $k_1,k_2\in[K].$ For any $e_{k_1,i}\in {\C U}_{k_1}$, there exists at most one vertex $e_{k_2,j}\in {\C U}_{k_2}$ such that $\left\{e_{k_1,i},e_{k_2,j}\right\}\in E({\C L}).$ 
\item[(C3)] 
For any $k$ and any vertex $e_{k,i}\in {\C U}_{k}$,  the set $\left\{e_{k,i}\cup {\C N}(e_{k,i})\backslash {\C U}_{k}\right\}$ containing $e_{k,i}$ and all adjacent vertices of $e_{k,i}$ except those in ${\C U}_{k}$, forms a clique. We refer to these cliques as the \underline{subfile-cliques}. Let $r$ be the number of subfile-cliques in $\C L$ and the subfile-cliques be denoted as ${\C S}_i:i\in[r]$. 
\item[(C4)]  The number of right vertices of $B$ is
\begin{equation}
\label{subpacket}
F=KD-\sum_{i=1}^r(|{\C S}_i|-1).
\end{equation}
\end{itemize}
\end{proposition}
\begin{IEEEproof}
We prove the If part. The Only If part can be inferred easily. We are given a graph $\C L$ satisfying properties (C1)-(C4). To prove the If part, we first create a bipartite graph $B_0$ with $K$ left vertices and $KD$ right vertices. Partitioning the right vertices into $K$ subsets of size $D$ each, we initialize the edge set of $B_0$ by assuming that the $k^{th}$ subset of right vertices in the partition are all adjacent to the $k^{th}$ left vertex. We also label the adjacent right vertices of $k$ as $k(i),i\in[D].$ Note that the line graph ${\C L}(B_0)$ contains $K$ cliques of size $D$ each and no other edges. Thus ${\C L}(B_0)$ is a subgraph of ${\C L}$ as (C1) holds. Furthermore, we note that by conditions (C1)-(C3), $E({\C L})=E({\C L}(B_0))\cup {\cal E}$, where ${\cal E}$ is the set of all edges in all the subfile cliques $S_j:j\in[r]$. 

The proof proceeds by updating the graph $B_0$ by identifying right-vertices according to the subfile-cliques of the given graph $\C L$ so that at any step $j$, the updated graph $B_j$ is such that ${\C L}(B_j)$ will be a subgraph of $B$. Finally we will have ${\C L}(B_r)={\C L}$, where $r$ is the number of subfile cliques. 

We proceed by induction. Assume that for $i\leq j-1$, ${\C L}(B_i)$ is a subgraph of ${\C L}$. We now explain how the graph $B_j$ is obtained from $B_{j-1}$, and show that ${\C L}(B_j)$ is a subgraph of ${\C L}$. Consider the $j^{th}$ subfile clique of $\C L$, given as ${\C S}_j=\left\{e_{k_{1},i_i},...,e_{k_{|{\C S}_j|},i_{|{\C S}_j|}}\right\}$.  To obtain $B_{j}$, we identify the right vertices $\left\{k_1(i_{1}),...,k_{|{\C S}_j|}\left(i_{|{\C S}_j|}\right)\right\}$ in $B_{j-1}$ as a single right vertex. Note that these vertices in $B_{j-1}$ are well-defined, as the subfile-cliques partition the vertices of the graph $\C L$ by Condition (C3). Furthermore, as (C2) holds, the vertices $k_i:i=1,...,|{\C S}_j|$ are all distinct. Hence $B_j$ continues to be $D$-left-regular, moreover with the same number of edges as $B_{j-1}$. With all these facts, we can see that $E({\C L}(B_j))=E({\C L}(B_{j-1}))\cup E',$ where $E'$ is the set of edges in the clique $S_j$. It follows that ${\C L}(B_j)$ is a subgraph of ${\C L}$. After the $r^{th}$ step, after going through all the subfile cliques, we have  $E({\C L}(B_r))=E({\C L})$ and hence ${\C L}(B_r)={\C L}$. 

We claim that the required bipartite graph is then $B=B_r$. It remains to check that the number of right vertices of $B$ is $KD-\sum_{i=1}^r(|{\C S}_i|-1)$. This is seen by noting that the number of right vertices of $B_0$ is $KD$, and the number of right vertices of $B_j$ is $(|{\C S}_j|-1)$ less than that of $B_{j-1}$, $\forall 1\leq j\leq r$. This completes the proof.
\end{IEEEproof}
By Proposition \ref{linegraph}, if we construct a graph ${\C L}$ satisfying conditions (C1)-(C3), then we have constructed a caching scheme based on a bipartite graph $B$ such that ${\C L}(B)={\C L}$ with subpacketization $F$ as in (\ref{subpacket}). We therefore give the following definition.
\begin{definition}
A graph ${\C L}$ is called a \underline{caching line graph} if it satisfies conditions (C1)-(C3) of Proposition \ref{linegraph} for some parameters $K$ and $D$. 
\end{definition}
Henceforth all our line graphs are caching line graphs. 
By Lemma \ref{coloringstrongedge}, any clique cover of $\Lbarsq$ (the complement of the square of $\C L$) gives us a transmission scheme (one transmission per clique) that satisfies all receiver demands. In order to obtain a clique cover of $\Lbarsq$, we have to understand the behaviour of the cliques of $\Lbarsq$. 
\begin{lemma}
\label{cliquelemma}
A subset of vertices ${\C C}\subset V(\Lbarsq)$ is a clique of $\Lbarsq$ if and only if the following condition is true.
\begin{itemize}
\item For any two vertices $e_{k_1,i_1},e_{k_2,i_2} \in {\C C}$, there exists no vertex in ${\C U}_{k_2}$ adjacent to  $e_{k_1,i_1}$ in $\C L$. 
\end{itemize}
Furthermore, any clique of ${\Lbarsq}$ contains at most one vertex from each of the user-cliques of $\C L$.
\end{lemma}
\begin{IEEEproof}
Recall that $V({\Lbarsq})=V({\C L}).$ We prove the only if part of the first claim. Let $\C C$ be a clique of $\Lbarsq$. 
Now, suppose there are vertices $e_{k_1,i_1},e_{k_2,i_2} \in {\C C}$ such that there is some vertex in ${\C U}_{k_2}$ adjacent to  $e_{k_1,i_1}$ in $\C L$. Then by definition of $\Lsq$, $e_{k_1,i_1}$ and $e_{k_2,i_2}$ will be adjacent in $\Lsq$, and hence non-adjacent in $\Lbarsq$. This contradicts our assumption that $\C C$ is a clique of $\Lbarsq$.

Now the if part. Suppose the condition of the lemma is satisfied for some subset ${\C C}\subset V(\C L)$, but there are two vertices $e_{k_1,i_1},e_{k_2,i_2} \in {\C C}$ which are non-adjacent in $\Lbarsq$, and hence adjacent in $\Lsq$. By definition of $\Lsq$, this means that either $\left\{e_{k_1,i_1},e_{k_2,i_2}\right\}\in E({\C L})$, or there exists some vertex $e_{k,i}$ such that $\left\{e_{k_1,i_1},e_{k,i}\right\}\in E({\C L})$ and $\left\{e_{k,i},e_{k_2,i_2}\right\}\in E({\C L})$. In the former case, there is a clear contradiction of the condition of the lemma statement. In the latter case, we have $\{e_{k_1,i_1},e_{k_2,i_2}\}\in {\C N}(e_{k,i})$. The case of $k_1=k_2$ is already handled by the former case. If no two of $\{k_1,k_2,k\}$ are equal to each other, this means $\{e_{k_1,i_1},e_{k_2,i_2},e_{k,i}\}$ must be in a clique of $\C L$ by (C3) of Proposition \ref{linegraph}, which violates the condition of lemma. Thus, WLOG, we can assume $k=k_2\neq k_1$, in which case we see that the vertex $e_{k,i}=e_{k_2,i}$ contradicts the assumed condition. This completes the proof of the first claim. 

The last claim of the lemma follows from the first claim, since if some clique of $\Lbarsq$ contains two vertices from the same user-clique of $\C L$ then the condition in the lemma is violated. 
\end{IEEEproof}
%
We now define a specific class of caching line graphs called $(c,d)$-caching line graphs. The reason for considering $(c,d)$-caching line graphs is because they yield easily to the computation of the rate and the subpacketization, as Theorem \ref{cliquecoverlinegraph} will show. 
\begin{definition}
A caching line graph $\C L$ such that $\C L$ has a clique cover consisting of $c$-sized disjoint subfile cliques and $\Lbarsq$ has a clique cover consisting of $d$-sized disjoint cliques, for some positive integers $c,d$, is called \underline{a $(c,d)$-caching line graph}. 
\end{definition}
\begin{theorem}
\label{cliquecoverlinegraph}
Consider a $(c,d)$-caching line graph ${\C L}$. Then there is a coded caching scheme consisting of the caching scheme given by ${\C L}$ with $F=\frac{KD}{c}$ (and thus $\frac{M}{N}=1-\frac{c}{K}$), and the associated   transmission scheme based on a clique cover of $\Lbarsq$ having rate $R=\frac{c}{d}$. Furthermore, if the number of files $N\geq K$, the rate $R$ of this scheme satisfies 
\begin{align}
\label{gaptooptimal}
R\leq \frac{R^*+\frac{1}{F}}{d\left(\frac{1}{F}+\frac{1}{K}\right)}.
\end{align}
where $R^*$ is the infimum of all achievable rates for $L$ with subpacketization $F=\frac{KD}{c}$.
\end{theorem}
\begin{IEEEproof}
Since there is a clique cover of ${\C L}$ (which satisfies (C1)-(C3)) with $c$-sized disjoint subfile cliques, by Proposition \ref{linegraph}, there exists a caching scheme with 
\[
F=KD-\sum_{i=1}^{\frac{KD}{c}}(c-1)=\frac{KD}{c}.
\]
Clearly, we also have $\frac{M}{N}=1-\frac{D}{F}=1-\frac{c}{K}.$

Since there is a clique cover of $\Lbarsq$ with $d$-sized cliques, by Lemma \ref{coloringstrongedge} and Section \ref{bipartite}, there exists a transmission scheme for the caching scheme defined by $\C L$, which consists of $\frac{KD}{d}$ transmissions, each transmission being a sum of $d$ subfiles. Thus the rate
\[
R=\frac{\frac{KD}{d}}{F}=\frac{c}{d}.
\]

Finally we should (\ref{gaptooptimal}). By Theorem \ref{thmratebound}, as $N\geq K$, we have
\begin{align}
\nonumber
R^*&\geq \left(\frac{K}{F}+1\right)\left(1-\frac{M}{N}\right)-\frac{1}{F}=\left(\frac{K}{F}+1\right)\frac{c}{K}-\frac{1}{F}\\
&\geq \frac{c-1}{F}+\frac{c}{K}=Rd\left(\frac{1}{F}+\frac{1}{K}\right)-\frac{1}{F}.
\end{align}
Thus we have proved
\[
R\leq \frac{R^*+\frac{1}{F}}{d\left(\frac{1}{F}+\frac{1}{K}\right)}.
\]
\end{IEEEproof}
\begin{remark}
\label{remarkcliquesize}
We observe that (\ref{gaptooptimal}) indicates that if the subpacketization $F$ is large compared to $K$ in a  bipartite caching scheme, then a clique cover of $\Lbarsq$ with cliques of size $\Theta(K)$ makes the rate $R$ of the transmission scheme based on the clique cover of $\Lbarsq$ close to the optimal rate $R^*$. Similarly if $K$ is much larger than $F$, a clique cover of $\Lbarsq$ with size $d$ being $\Theta(F)$ brings $R$ close to optimal.
\end{remark}
In the rest of this section, we reinterpret some priorly known coded caching schemes as schemes based on caching line graphs. In both these examples, it may be observed that the situation is similar to that of Remark \ref{remarkcliquesize};  we have $F$ growing exponentially in $K$ as $K\rightarrow \infty$,  but $d=\Theta(K)$ and hence we can keep the rate close to optimal.
\begin{example}
\label{exmcaching1}
For given parameters $M,K,N$, let $t=\frac{MK}{N}$. We will now construct a $(K-t,t+1)$-caching line graph, which corresponds to the coded caching scheme of \cite{MaN}. The caching line  graph ${\C L}$ is initialized with $K$ cliques of size $D=\binom{K-1}{t}$, indexed using $[K]$. For each user $i\in[K]$, denote the $D$ vertices of the $i^{th}$ user-clique as $\{(i,A):A\subset [K]\backslash i, |A|=t\}$. 

For each $A\subseteq [K]$ such that $|A|=t$, we create a clique ${\cal C}_A$ of size $K-t$ in $\C L$ consisting of the vertices $\{(i,A):i\in[K],i\notin A\}$ by defining edges between all these vertices. It is easy to see that 
\[
\bigcup\limits_{A\subset[K]:|A|=t} {\C C}_A = V({\C L}).
\]

For some $(t+1)$-sized $B\subset [K]$, consider the set of vertices of ${\C L}$ given by
\[
{\C C}'_B=\{(i,B\backslash i):i\in B\}
\]
 consisting of $t+1$ vertices of $\C L$. It is not difficult to see that for any distinct $i_1,i_2\in B$, there exists no edge in $\C L$ from the vertex $(i_1,B\backslash i_1)$ to any vertex in the $i_2^{th}$ user-clique. Thus, by Lemma \ref{cliquelemma}, ${\C C}'_B$ forms a clique in $\Lbarsq$ of size $(t+1)$. Also note that 
 \[
\bigcup\limits_{B\subset[K]:|B|=t+1} {\C C}'_B = V({\C L})=V({\Lbarsq}).
\]
Thus, the caching line graph $\C L$ is a $(K-t,t+1)$-caching line graph. Hence by Theorem \ref{cliquecoverlinegraph}, the subpacketization for this graph is 
\[
F=\frac{KD}{K-t}=\frac{K}{K-t}\binom{K-1}{t}=\binom{K}{t}.
\] 
And the rate corresponding to the clique cover scheme on $\Lbarsq$ is 
\[
R=\frac{K-t}{t+1}=\frac{K(1-\frac{M}{N})}{\frac{MK}{N}+1}.
\]
We have thus recovered the coded caching scheme of \cite{MaN} using $\C L$.
\end{example}
\begin{example}
\label{exmcaching2}
We now recover a special case of the coded caching scheme based on resolvable deisgns  from \cite{TaR} which first appeared in \cite{TaR1}. Let ${\mathfrak C}$ be a $(k-1)$ dimensional linear single parity check code of length $k$  over a finite field ${\mathbb F}_q$.  We initialize the caching line graph $\C L$ with $K=kq$ user-cliques, each consisting of  $D=q^{k-1}-q^{k-2}$ vertices. We index the user-cliques as $U_{i,l}$, where $i\in [k]$, and $l\in {\mathbb F}_q$. The vertices of the user-clique $U_{i,l}$ are indexed as follows. 
\[
U_{i,l}=\{(\boldsymbol{v},i,l):\boldsymbol{v}=(v_1,\hdots,v_k)\in{\mathfrak C}~\text{and}~ v_i\neq l\}.
\]
It is not difficult to see that  
\[
|{\mathfrak C}\backslash U_{i,l}|=|\{\boldsymbol{v}=(v_1,\hdots,v_k)\in{\mathfrak C}:v_i=l\}|=q^{k-2},
\]
since we can think of ${\mathfrak C}\backslash U_{i,l}$ as a coset of the subcode ${\mathfrak C}\backslash U_{i,0}$ within $\mathfrak C$. For a formal proof, we refer the reader to \cite{TaR1}.  Thus,  $|U_{i,l}|=|{\mathfrak C}|-|{\mathfrak C}\backslash U_{i,l}|=q^{k-1}-q^{k-2}=D$.

We now construct the subfile cliques as follows. For each vector $\boldsymbol{v}\in {\mathfrak C}$, we construct the clique 
\[
{\cal C}_{\boldsymbol{v}}=\{(\boldsymbol{v},i,l): \forall i\in[K], \forall l\in{\mathbb F}_q\}
\] 
by creating the edges between all the vertices in ${\cal C}_{\boldsymbol v}$.
Again, it is not difficult to see that $\cup_{\boldsymbol{v}\in{\mathfrak C}}{\cal C}_{\boldsymbol{v}}=V({\C L}).$ Thus the cliques $\{{\C C}_{\boldsymbol{v}}:\boldsymbol{v}\in {\mathfrak C}\}$ form a disjoint clique cover of $\C L$. Furthermore $|{\C C}_{\boldsymbol{v}}|=kq-k$, since by definition, an user-clique $U_{i,l}$ does not contain $(\boldsymbol{v},i,l)$ if and only if $v_i=l$ and thus $|\{U_{i,l}:\forall i,l \text{ s.t } (\boldsymbol{v},i,l)\notin U_{i,l}\}|=k$. 

From the above construction of the subfile-clique cover for $\C L$ we have from Theorem \ref{cliquecoverlinegraph} that $F=\frac{KD}{k(q-1)}=q^{k-1}$. We now construct a clique cover of $\Lbarsq$. For ${\B l}=(l_1,\hdots,l_k)\in{\mathbb F}_q^k\backslash{\mathfrak C}$,  let 
$
{\B l}(i)$ be the codeword in $\mathfrak C$ such that $
{\B l}(i)$  is equal to  $\B l$ at the coordinates $[k]\backslash i$ but not at the $i^{th}$ coordinate. Note that a unique such codeword does exist in $\mathfrak C$ by definition of $\mathfrak C$ and ${\B l}$. Now consider the set of vertices of $\C L$ given by
\[
{\C C}'_{\B l}=\left\{({\B l}(i),i,l_i)~:~i\in[k]\right\}.
\]
Note that $({\B l}(i),i,l_i)\in U_{i,l_i}, \forall i\in[k]$. Also, for $i\neq j$, there exists no edge from $({\B l}(i),i,l_i)$ to any vertex in $U_{j,l_{j}}$  because in ${\B l}(i)$, the $j^{th}$ coordinate is precisely $l_j$. Thus ${\C C}'_{\B l}$ forms a clique of size $k$ in $\Lbarsq$. Furthermore, it is not hard to see that 
\[
\bigcup_{{\B l}\in{\mathbb F}_q^k\backslash{\mathfrak C} }{\C C}'_{\B l}=V({\C L})=V({\Lbarsq}),
\]
where the above union is a disjoint union. Thus the $k$-sized disjoint cliques ${\C C}'_{\B l}$s cover the vertices of $\Lbarsq$. We have thus got a $(kq-k,k)$-caching line graph $\C L$. By Theorem \ref{cliquecoverlinegraph}, the coded caching scheme on $\C L$  has rate $\frac{k(q-1)}{k}=q-1$. We have hence recovered the scheme from \cite{TaR1}.
\end{example}
\begin{remark}
In the examples given so far in this section, we have essentially reverse engineered the schemes given in prior works and demonstrated how they can be intrepreted according to the line graph framework we have presented in this current work. We also remark that the caching schemes based on PDAs (placement delivery arrays) discussed in \cite{YCTC} and subsequent works can be seen in the framework of caching line graphs as well. However the special class of $(c,d)$-caching line graphs seem to offer some advantages in terms of tracking the subpacketization and rate using the graph characteristics. In the forthcoming section, we present a new explicit construction of a caching scheme based on $(c,d)$-caching line graphs.
\end{remark}
\section{A Line Graph based Coded Caching Scheme based on Projective Geometry}
\label{ourscheme}
We recollect some basic ideas of projective geometries over finite fields. The reader is referred to \cite{Hirsch} for more details. For positive integers $k$, let $PG_q(k-1)$ denote  the $(k-1)$-dimensional projective space over $\fq$. The elements of $PG(k-1,q)$ are called the \textit{points} of $PG_q(k-1)$. The points of $PG_q(k-1)$ can be considered as  the representative vectors of one-dimensional subspaces of $\fq^k$. For $m\geq 1, 1\leq m\leq k$, let $PG_q(k-1,m-1)$ denote the set of $m$-dimensional subspaces of $\fq^k$. It is known that $|PG_q(k-1,m-1)|$ is equal to the Gaussian binomial coefficient $\gbinom{k}{m}$, where $\gbinom{k}{m}$ is given by 
\[
\begin{bmatrix}k\\m\end{bmatrix}_q
=\frac{(q^k-1)\hdots(q^{k-m+1}-1)}{(q^m-1)\hdots(q-1)}.
\]
In any Gaussian binomial coefficient $\gbinom{a}{b}$ given in this paper we assume that $1\leq b\leq a.$
The following is known about the Gaussian binomial coefficients (see \cite{Hirsch}, for example).
\begin{lemma}
\label{gaussiancoeff}
~\\
\begin{itemize}
\item The Gaussian binomial coefficient $\gbinom{k}{m}$ is the number of subspaces of dimension $m$ of any $k$-dimensional subspace over $\fq$. Also, $\gbinom{k}{m}=\gbinom{k}{k-m}$.
\item The number of elements of $PG_q(k-1,m-1)$ that contain a given $t$-dimensional subspace ($1\leq t\leq m$) is 
\[
\frac{(q^{k-t}-1)\hdots(q^{k-m+1}-1)}{(q^{m-t}-1)\hdots(q-1)}=\gbinom{k-t}{m-t}.
\]
\end{itemize}
\end{lemma}
In the following subsection, we give a construction of a coded caching scheme based on projective geometry. The scheme we present can be thought of a $q$-analogue of a generalization of the original scheme of \cite{MaN}. As in Examples \ref{exmcaching1} and \ref{exmcaching2}, we first give the caching line graph $\C L$ by describing its user-cliques and subfile-cliques (each of same size), and then show that there is a clique cover of $\Lbarsq$ containing cliques of the same size. 
\subsection{Coded Caching Scheme Construction}
\label{construction}
Consider positive integers $k,m,t$ such that $m+t\leq k$. Let $K=\gbinom{k}{t}$. We first initialize $\C L$ by its user-cliques. The user-cliques are identified by $t$-dimensional subspaces of $\fq^k$. For each $t$-dimensional subspace $V$ of $\fq^k$, create the vertices corresponding to the user-clique identified by $V$,
\[
{\C C}_{V}=\{(V,X),\forall X\in PG_q(k-1,m+t-1): V\subseteq X\}.
\]
Thus, $D=|{\C C}_V|=\gbinom{k-t}{(m+t)-t}=\gbinom{k-t}{m}$ by Lemma \ref{gaussiancoeff}. For each $(m+t)$-dimensional subspace $X$ of $\fq^k$, we construct the subfile clique of $\C L$  associated with $X$ as 
\[
{\C C}_X=\{(V,X)\in V({\C L}): \forall V~\text{such that}~V\subseteq X\}.
\]
It's not difficult to see that the cliques $\{{\C C}_X: X\in PG_q(k-1,m+t-1)\}$ partition $V(\C L)$. 

In order to decide on the transmission scheme, we have to obtain a clique cover of $\Lbarsq$.  The clique cover of $\Lbarsq$ that we wish to obtain is based on a relabelling of the vertices of $\C L$ based on $m$-dimensional subspaces of $\fq^k$. Towards that end, we first require the following lemmas(Lemma \ref{perfectmatching} and Lemma \ref{validalternatelabeling}) using which we can find `matching' labels to the $t$-dimensional and $m$-dimensional subspaces of some $X \in PG_q(k-1,m+t-1)$. Subsequently, using Lemma \ref{cliqueoflbarsqproj} and Lemma \ref{cliquecoveroflbarsqproj}, we show the clique cover of $\overline{{\C L}^2}$.
 \begin{lemma}
\label{perfectmatching}
Consider some element $X\in PG_q(k-1,m+t-1)$. Let $\left\{ V_i, i=1,\hdots,\gbinom{m+t}{t}\right\}$ denote the $t$-dimensional subspaces of $X$ taken in some fixed order.  Then the set of $m$-dimensional subspaces of $X$ can be written as an ordered set as $\left\{T_i,~ i=1,\hdots,\gbinom{m+t}{m}\right\}$ such that $T_i \oplus V_i=X, \forall i$ (where $\oplus$ denotes direct sum). Moreover such an ordering can be found in operations polynomial in $\gbinom{m+t}{t}$.
\end{lemma}
\begin{IEEEproof}
See Appendix \ref{appendixproofperfectmatching}.
\end{IEEEproof}
For a $t$-dimensional space $V_i$ contained in  a $(m+t)$-dimensional space $X,$ let $T_i$ (the $m$-dimensional subspace as obtained in Lemma \ref{perfectmatching} such that $T_i\oplus V_i=X$) be called \textit{the matching subspace of $V_i$ in $X$}. Using these matching subspaces, we can obtain an alternate labeling scheme for the vertices of our caching line graph ${\C L}$. The alternate labels are given as follows.
\begin{itemize}
\item Let the alternate label for $(V,X)$ be $(V,T_{V,X})$, where $T_{V,X}$ is the $m$-dimensional matching subspace of $V$ in $X$ obtained using Lemma \ref{perfectmatching}.
\end{itemize}
The following lemma ensures that the alternative labeling given above is indeed a valid labelling, i.e., it uniquely identifies the vertices of ${{\C L}}$. 
\begin{lemma}
\label{validalternatelabeling}
No two vertices of $V({{\C L}})$ have the same alternate label, i.e., if $(V_1,X_1),(V_2,X_2)\in V({{\C L}})$ have the same alternate label $(V,T_{V,X})$, then $(V_1,X_1)=(V_2,X_2)$. 
\end{lemma}
\begin{IEEEproof}
If $(V_1,X_1),(V_2,X_2)\in V({{\C L}})$ have the same alternate label $(V,T_{V,X})$, then clearly $V_1=V_2=V$. Moreover we should also, by definition of the alternate labels, have that $X_1=T_{V,X}\oplus V=X_2$. Hence proved.
\end{IEEEproof}
We are now in a position to present the clique-cover of ${\overline{{\C L}^2}}$. Our cliques are represented in terms of the alternate labels given to the vertices of ${\C L}$. We first show the structure of one such clique.
\begin{lemma}
\label{cliqueoflbarsqproj}
For a $m$-dimensional subspace $T$ of $\fq^k$, consider the set of vertices of $\overline{{\C L}^2}$ (identified by their alternate labels) as follows. 
\[
{\cal C}_T=\{(V,T)\in V({\C L}): V\in PG_q(k-1,t-1)\}.
\]
Then ${\C C}_T$ is a $\gbinom{k-m}{t}$-sized clique of $\overline{{\C L}^2}$. 
\end{lemma}
\begin{IEEEproof}
Firstly, we observe that ${\C C}_T$ is a well-defined set because the $T$ is an $m$-dimensional subspace of precisely $\gbinom{k-m}{(m+t)-m}$  $(m+t)$-dimensional subspaces by Lemma \ref{gaussiancoeff}. 


Note that $(V,T)$ is the alternate label for $(V,T\oplus V)\in {\C C}_{V}$ (the user-clique indexed by $V$). Also we can observe that for distinct $(V_1,T), (V_2,T)\in {\C C}_T$, we must have $V_1\oplus T\neq V_2\oplus T$. This is due to the fact that each $m$-dimensional subspace within a $(m+t)$-dimensional subspace $X$ is matched to a unique $t$-dimensional subspace of $X$. Hence, by Lemma \ref{perfectmatching} and our alternate labeling scheme, we should have $|{\C C}_T|=\gbinom{k-m}{(m+t)-m}=\gbinom{k-m}{t}$. 

We now show that for any distinct $(V_1,T),(V_2,T)\in {\C C}_T,$ there exists no edge in ${\C L}$ between  $(V_1,T)$ and  any vertex in ${\C C}_{V_2}$. Invoking Lemma \ref{cliquelemma} completes the proof. 

Note that by our construction of ${\C L}$, suppose an edge exists in ${\C L}$  between $(V_1,T)$ and some vertex (say $(V_2,T')$) in ${\C C}_{V_2}$, then the vertices $(V_1,T),(V_2,T')$ would be part of the subfile clique ${\C C}_{T\oplus V_1}$. Thus, we would have $V_2\subset T\oplus V_1$. This would mean that  $T\oplus V_1= T\oplus V_2$, which is a contradiction. 
This completes the proof. 
\end{IEEEproof}
We now show that the cliques $\left\{{\C C}_T~:~T\in PG_q(k-1,m-1)\right\}$ partition $V({{\C L}})$. 
\begin{lemma}
\label{cliquecoveroflbarsqproj}
\[
\bigcup\limits_{T\in PG_q(k-1,m-1)}{\C C}_T=V({{\C L}}),
\]
where the above union is a disjoint union.
\end{lemma}
\begin{IEEEproof}
It should be clear from our alternate labeling scheme and the definition of ${\C C}_{T}$ that any vertex $(V,X)\in {\C C}_V$ (which gets some alternate label $(V,T_{V,X})$) appears at least in one clique of $\Lbarsq$, i.e., ${\C C}_{T_{V,X}}.$ Furthermore, by definition ${\C C}_{T_1}$ and ${\C C}_{T_2}$ are disjoint for any two $T_1$ and $T_2$. This completes the proof.
\end{IEEEproof}
\begin{theorem}
\label{projlinegraphparameters}
The caching line graph ${\C L}$ constructed in Section \ref{construction} is a $\left(\gbinom{m+t}{t},\gbinom{k-m}{t}\right)$-caching line graph and defines a coded caching scheme with $K=\gbinom{k}{t}$, $ F=\gbinom{k}{m+t}$, $\frac{M}{N}=1-\frac{\gbinom{m+t}{t}}{\gbinom{k}{t}}$, and $R=\frac{\gbinom{m+t}{t}}{\gbinom{k-m}{t}}.$
\end{theorem}
\begin{IEEEproof}
By our construction, $K=\gbinom{k}{t}$. For any $X\in PG_q(k-1,m+t-1)$, the size of the subfile-clique $|{\C C}_X|=\gbinom{m+t}{t}$. The size of each user-clique $D=\gbinom{k-t}{m}$.  By our construction, the size of the cliques of  $\Lbarsq$ is $\gbinom{k-m}{t}$ and they partition the vertices. Hence  ${\C L}$ is a  $\left(\gbinom{m+t}{t},\gbinom{k-m}{t}\right)$-caching line graph.
Thus, we have by Theorem \ref{cliquecoverlinegraph},
\[
F=\frac{\gbinom{k}{t}\gbinom{k-t}{m}}{\gbinom{m+t}{t}}=\gbinom{k}{m+t}.
\]
 And also, by Theorem \ref{cliquecoverlinegraph}, we have $\frac{M}{N}=1-\frac{\gbinom{m+t}{t}}{\gbinom{k}{t}}.$ The rate calculation follows similarly. 
\end{IEEEproof}
\section{Analysis of the scheme}
\label{analysis}
In this section, we analyse the coded caching scheme in Section \ref{ourscheme}. We are essentially interested in finding out some asymptotic results about the scheme. For this reason, we use the following simple upper and lower bounds on Gaussian binomial coefficients and their relationships. 
\begin{lemma}
\label{approximations}
For non-negative integers $a,b,f$, for $q$ being some prime power, 
\begin{align}
\label{eqn31}
&q^{(a-b)b}&\leq &\gbinom{a}{b} \leq & q^{(a-b+1)b}\\
\label{eqn32}
&q^{(a-f-1)b} &\leq &\frac{\gbinom{a}{b}}{\gbinom{f}{b}}\leq & q^{(a-f+1)b}\\
\label{eqn33}
&q^{(a-f-b-1)\delta} &\leq &\frac{\gbinom{a}{b}}{\gbinom{a}{f}} \leq & q^{(a-f-b+1)\delta},
\end{align}
where $\delta=max(b,f)-min(b,f)$. 
\end{lemma}
\begin{IEEEproof}
The first lower bound for $\gbinom{a}{b}$ is well known from combinatorics literature (see for instance, \cite{Tak}). All the other bounds are proved by definition of the Gaussian binomial coefficient and by noting that $q^a-1\geq q^{a-1}$ (since $q\geq 2$), and $q^a-1\leq q^a$. 
\end{IEEEproof}
Using the above bounds, we provide a simple analysis of our scheme. Throughout we assume $q$ is constant. We have $K=\gbinom{k}{t}$. We analyse our scheme as $k$ grows large. 
Consider 
\begin{align}
\label{eqn34}
1-\frac{M}{N}=\frac{\gbinom{m+t}{t}}{\gbinom{k}{t}} \stackrel{(\ref{eqn32})}{\geq} q^{(m+t-k-1)t}.
\end{align}
Suppose the choice of $t,m$ are such that $q^{(m+t-k-1)t}$ is constant, i.e., $t$ and $k-m$ are constants for this purpose. We then have the following. 
\begin{align}
\nonumber
F=\gbinom{k}{m+t}&\stackrel{(\ref{eqn33})}{\leq}q^{(k-t-m-t+1)m}\gbinom{k}{t}\\
\label{eqn35}
&\leq q^{(k-t)(k-2t-m+1)}\gbinom{k}{t} (\text{since}~ k\geq m+t)\\
\nonumber
&\leq K^{\frac{k-2t-m+1}{t}}K, (\text{since }K\geq q^{t(k-t)}~\text{by}~(\ref{eqn31}))\\
&\leq K^{\frac{k-t-m+1}{t}}
\end{align} 
which is clearly $O(poly(K))$ because of our choice of constants. 

We have to still determine the behavior of the rate with respect to our choice of parameters. We have,
\begin{align*}
R&=\frac{\gbinom{m+t}{t}}{\gbinom{k-m}{t}}\stackrel{(\ref{eqn31})}{\geq} \frac{K(1-\frac{M}{N})}{q^{(k-m-t+1)t}}\stackrel{(\ref{eqn34})}{\geq} \frac{Kq^{(m+t-k-1)t}}{q^{(k-m-t+1)t}}\\
&\geq \frac{K}{q^{2(k-m-t+1)t}}.\end{align*}
By a similar process, we can obtain
\begin{align*}
R\leq \frac{K}{q^{2(k-m-t-1)t}}.
\end{align*}
Thus it is clear that $R=\Theta(K)$, similar to the uncoded case, due to our choice of constants. 

On the other hand, suppose we want to make the rate upper bounded by a constant. We have
\begin{align*}
R\stackrel{(\ref{eqn32})}{\leq} q^{(m+t-k+m+1)t}.
\end{align*} 
Thus we keep $t$ and $k-2m$ as constants and continue.

We once again have by (\ref{eqn35})	
\begin{align*}
F&\leq q^{(k-2t-m+1)(k-t)}K=q^{(k-t)(\frac{k}{2}+\frac{k-2m}{2}-2t+1)}K\\
&\leq q^{(\frac{1}{t}log_q K)((\frac{1}{2t}log_q K+\frac{t}{2})+\frac{k-2m}{2}-2t+1)}q^{log_qK},\\
\end{align*}
where the last inequality follows since $K\geq q^{t(k-t)}$ by $(\ref{eqn31})$. Hence we have $F=q^{O((log_qK)^2)}$.  Finally we focus on the uncached fraction $1-\frac{M}{N}$. We have by (\ref{eqn34}),
\begin{align*}
1-\frac{M}{N} &\geq q^{(m+t-k-1)t}=q^{(m-\frac{k}{2}-\frac{k}{2}+t-1)t}\\
&\geq q^{(m-\frac{k}{2}+t-1)t}\frac{1}{\sqrt{K}}q^{-\frac{t^2}{2}}~(\text{since }K\geq q^{t(k-t)}~\text{by}~(\ref{eqn31}))
\end{align*}
Thus we have $1-\frac{M}{N}\geq \frac{1}{\sqrt{K}}q^{(m-\frac{(k-t)}{2}-1)t}.$ By a similar technique, we can obtain that $1-\frac{M}{N}\leq \frac{1}{\sqrt{K}}q^{(m-\frac{(k-t)}{2}+1)t}.$ Hence $1-\frac{M}{N}=\Theta(\frac{1}{\sqrt{K}})$.

In Table \ref{tab3}, we also give a comparison of the actual quantities between our scheme and that of \cite{MaN} for some particular choice of users and the uncached fraction (as defined by our scheme). 
\begin{table}
\centering
\begin{tabular}{|c|c||c|c||c|c|}
\hline
$K$ & $1-\frac{M}{N}$ &$F$ &  $F$ \cite{MaN} & $R$  & $R$ \cite{MaN}\\
&&\scriptsize (this work)&& \scriptsize (this work)&\\
$\tiny \gbinom{k}{t}$ & $\tiny \frac{\gbinom{m+t}{t}}{\gbinom{k}{t}}$ & $\tiny \gbinom{k}{m+t}$ & $\tiny \binom{K}{\frac{KM}{N}}$  & $\tiny \frac{\gbinom{m+t}{t}}{\gbinom{k-m}{t}}$ & $\tiny \frac{K(1-\frac{M}{N})}{\frac{MK}{N}+1}$\\
\hline
($k=6$ & $(m=3)$ &&&&\\
$t=2$) &$\frac{5}{21}$&63&$\binom{651}{496}$ & $\frac{155}{7}$ & $\frac{155}{497}$\\
$651$ &  & & &&\\
\hline
($k=6$ & ($m=3$) &&&&\\
$t=1$) &$\frac{5}{21}$ & 651 & $\binom{63}{48}$ & $\frac{15}{7}$ & $\frac{15}{49}$\\
$63$ & &&&&\\
\hline
($k=8$ & ($m=6$) &&&&\\
$t=1$) &$\frac{1}{17}$ & \scriptsize 200787 & $\binom{255}{240}$ & $\frac{15}{31}$ & $\frac{15}{241}$\\
$255$ & &&&&\\
\hline
($k=8$ & ($m=4$) &&&&\\
$t=1$) &$\frac{31}{197}$ & 2667 & $\binom{127}{96}$ & $\frac{31}{7}$ & $\frac{31}{97}$\\
$127$ & &&&&\\
\hline
\end{tabular}
\caption{For some specific values of $K,1-\frac{M}{N}$, we compare the results of \cite{MaN} with this work.}
\label{tab3}
\end{table}
\section{Conclusion}
\label{conclusion}
In this work, we have presented a framework for constructing coded caching schemes for broadcast networks via line graphs of bipartite graphs, building on results from \cite{YTCC}. Firstly a subpacketization-dependent lower bound on the rate is  derived using the bipartite graph framework for caching. The existence of $(c,d)$-caching line graphs enables us to nicely characterize the three important quantities, fractional cache requirement ($\frac{M}{N}$), the subpacketization level ($F$) and the rate $R$ based on graph theoretic parameters. We then present one explicit construction of such a $(c,d)$-caching line graph using projective geometry.  For the uncached fraction $(1-\frac{M}{N}$) lower bounded by a constant, this scheme achieves subpacketization $F=O(poly(K))$, however rate $R$ is $O(K)$. In another regime of operation where the rate remains below a constant, we get $F=q^{O(log_qK)^2}$ while the uncached fraction $1-\frac{M}{N}$ is $O(\frac{1}{\sqrt{K}})$.  Unfortunately it appears that the scheme in this paper can hold only one parameter (among $R,\frac{M}{N},F$) bounded by a constant, with the other two vary with $K$. Other schemes based on $(c,d)$-caching line graphs could prove to be useful in arriving at coded caching schemes with more interesting and useful parameters.
\section*{Acknowledgment} The author would like to thank Dr. Girish Varma for fruitful discussions regarding this work. 

\appendices
\section{Proof of Lemma \ref{perfectmatching}}
\label{appendixproofperfectmatching}
Construct a bipartite graph with left vertices as $\left\{ V_i, i=1,\hdots,\gbinom{m+t}{t}\right\}$ and right vertices as $\left\{T	 : T~\text{is a}~m-\text{dimensional subspace of}~X\right\}$. By Lemma \ref{gaussiancoeff}, the number of right-vertices is  $\gbinom{m+t}{m}=\gbinom{m+t}{t}$, the number of left vertices. For a left vertex $V$, let the adjacent right-vertices in the bipartite graph be $\{T: V\cap T=\phi\}$. Thus the left-degree is $\gbinom{m+t}{t}-|\{T: T\cap V\neq \phi\}|$. Now, $T\cap V$ is a subspace. It is known (for instance, see Chapter 3, Theorem 3.3, in \cite{Hirsch}) that the number of $m$-dimensional subspaces of $X$ which intersect with a given $t$-dimensional subspace in some subspace of dimension $i$ ($1\leq i\leq min(t,m)$) is 
\begin{align*}
q^{((m-1)-(i-1))((t-1)-(i-1))}\gbinom{(m+t-1)-(t-1)}{(m-1)-(i-1)}\gbinom{t}{i} \\
= q^{(m-i)(t-i)}\gbinom{m}{i}\gbinom{t}{i}.
\end{align*}
Thus the left-degree in this bipartite graph is 
\[
\gbinom{m+t}{t}-\sum_{i=1}^{min(m,t)}q^{(m-i)(t-i)}\gbinom{m}{i}\gbinom{t}{i},
\]
where the second term above is precisely $|\{T: T\cap V\neq \phi\}|$.

Similarly, the number of $t$-dimensional subspaces which intersect with a given $m$-dimensional subspace in some subspace of dimension $i$ is known \cite{Hirsch} as
\begin{align}
q^{(t-i)(m-i)}\gbinom{(m+t-1)-(m-1)}{(t-1)-(i-1)}\gbinom{m}{i}\\=q^{(t-i)(m-i)}\gbinom{t}{i}\gbinom{m}{i}.
\end{align}
And hence the right degree is equal to the left-degree.
%
%
Hence the bipartite graph we have constructed is regular. 

A perfect matching of a graph $G$ is a matching of $G$ such that every vertex of $G$ is incident on some edge of the matching. It should be clear that what we are looking for is a perfect matching of the regular bipartite graph we have constructed. The reason is as follows. Let $T_i$ be the $m$-dimensional subspace  adjacent to $V_i$ in the perfect matching. Since for given $V_i$, any $T$ adjacent to $V_i$ in our bipartite graph is such that $T \oplus V_i = X$, thus we have $T_i\oplus V_i=X$. Thus the ordering of the right-vertices that we desire as per the lemma statement can be obtained from the perfect matching. 
 
Now, for a regular bipartite graph with $n$ left-vertices, algorithms are known to find a perfect matching with complexity as small as $O(nlog n)$ \cite{GKK}. This completes the proof.

\begin{thebibliography}{160}
\bibitem{Cis}``Cisco visual networking index: Global mobile data traffic forecast update, 2015-2020'', [Online]. Available: http://goo.gl/1XYhqY.
\bibitem{MaN}
M.A. Maddah-Ali and U. Niesen, ``Fundamental limits of caching'', IEEE Transactions on Information Theory, Vol. 60, No. 5, Mar 2014, pp. 2856-2867.
%
\bibitem{JCM}
M. Ji, G. Caire, and A. F. Molisch,``Fundamental limits of caching in wireless d2d networks'', IEEE Trans. Inform. Theory, vol. 62, no. 2, pp. 849-869, Feb. 2016.
\bibitem{PMN}
R. Pedarsani, M. A. Maddah-Ali, and U. Niesen, ``Online Coded Caching'', IEEE/ACM Trans. on Networking, Vol. 24, No. 2, April 2016, pp. 836 - 845.  
\bibitem{YMA}
Q. Yu, M.A. Maddah-Ali and A.S. Avestimehr, ``The exact rate-memory tradeoff for caching with uncoded prefetching'', IEEE International Symposium on Information Theory, 2017, held at Aachen, Germany, 25-30 June, pp. 1613-1617.
%
\bibitem{SJTLD}
K. Shanmugam, M. Ji, A. M. Tulino, J. Llorca, and A. G. Dimakis, ``Finite-length analysis of caching-aided coded multicasting'', IEEE Trans. Inf. Theory, Vol. 62, No. 10, pp. 5524-5537, Oct. 2016.
\bibitem{YTCC}
Q. Yan, X. Tang, Q. Chen, M. Cheng, ``Placement Delivery Array Design through Strong Edge Coloring of Bipartite Graphs'', IEEE Comm. Letters, Vol. 22, Issue 2, Feb. 2018, pp. 236-239.
%
\bibitem{YCTC}
Q. Yan, M. Cheng, X. Tang, Q. Chen, ``On the Placement Delivery Array Design for Centralized Coded Caching Scheme'', IEEE Trans. on Info. Theory, Vol. 63, No. 9, Sep. 2017, pp. 5821-5833.
\bibitem{SZG}
C. Shangguan, Y. Zhang and G. Ge, ``Centralized coded caching schemes: A hypergraph theoretical approach'', Available at https://arxiv.org/abs/1608.03989, Aug. 2016.
\bibitem{STD}
K. Shanmugam, A. M. Tulino, and A. G. Dimakis, ``Coded Caching with Linear Subpacketization is Possible using Ruzsa-Szeméredi Graphs'', IEEE ISIT 2017, 
\bibitem{TaR}
L. Tang, A. Ramamoorthy, ``Coded Caching Schemes with Reduced Subpacketization from Linear Block Codes'', IEEE Trans. on Info. Theory, Vol. 64, No. 4, April 2018, pp. 3099-3120. 
%
\bibitem{TaR1}
 L. Tang, A. Ramamoorthy, "Coded caching for networks with the resolvability property", Proc. IEEE Int. Symp. Inf. Theory, pp. 420-424, Jul. 2016.
\bibitem{CYTJ}
M. Cheng, Q. Yan, X. Tang, J. Jiang, ``Coded Caching Schemes with Low Rate and Subpacketizations'', Available at https://arxiv.org/abs/1703.01548, Oct 2017.
%
\bibitem{CJYT}
M. Cheng,  J. Jiang, Q. Yan, X. Tang, ``Coded Caching Schemes for Flexible Memory Sizes'', Available at https://arxiv.org/abs/1708.06650, Oct 2017.
\bibitem{SLB}
S. A. Saberali, L. Lampe, I. Blake, ``Decentralized Coded Caching Without File Splitting'', Available at http://arxiv.org/abs/1708.07493v1, Aug 2017.
%
\bibitem{Die}
R. Diestel, ``Graph Theory'', Second Edition, Springer-Verlag, 2000. 
%
%
\bibitem{Hirsch}
J. W. P. Hirschfeld, ``Projective Geometries over Finite Fields'', Second Edition, Oxford Mathematical Monographs, Clarendon Press, 1998. 
%
\bibitem{GKK}
A. Goel, M. Kapralov, and S. Khanna, ``Perfect Matchings in O(n log n) Time in Regular Bipartite Graphs'',  Proceedings of the 42nd ACM Symposium on Theory of Computing (STOC), 2010, SIAM J. Comput., 42(3), pp. 1392–1404. 
\bibitem{Tak}
L. Takacs, ``Some asymptotic formulas for lattice paths'', Journal of Statistical Planning and Inference, Vol. 14, 1986, pp. 123-142. 
\end{thebibliography}
\end{document}